\documentclass[aps,prl,reprint,floatfix]{revtex4-1}
\usepackage[T1]{fontenc}

\usepackage{hyperref}
\usepackage{graphicx}
\usepackage{float}
\usepackage[caption = false]{subfig}
\usepackage{multirow}
\usepackage{color}
\usepackage{amsmath,amsfonts,amssymb}
\let\oldAA\AA
\renewcommand{\AA}{\text{\normalfont\oldAA}}

\newcommand{\xone }{$\times$}

\begin{document}

\title{Competition between Ta-Ta and Te-Te bonding leading to the commensurate charge density wave in  \texorpdfstring{TaTe$_4$}{TaTe4}}

\author{Bogdan Guster}
\affiliation{Catalan Institute of Nanoscience and Nanotechnology (ICN2), CSIC and The Barcelona Institute of Science and Technology, Campus Bellaterra, 08193 Barcelona, Spain}
\affiliation{European Theoretical Spectroscopy Facility, Institute of Condensed Matter and Nanosciences, Universit\'{e} catholique de Louvain, Chemin des \'{e}toiles 8, bte L07.03.01, B-1348 Louvain-la-Neuve, Belgium}

\author{Miguel Pruneda}
\affiliation{Catalan Institute of Nanoscience and Nanotechnology (ICN2), CSIC and The Barcelona Institute of Science and Technology, Campus Bellaterra, 08193 Barcelona, Spain}

\author{Pablo Ordej\'on}
\affiliation{Catalan Institute of Nanoscience and Nanotechnology (ICN2), CSIC and The Barcelona Institute of Science and Technology, Campus Bellaterra, 08193 Barcelona, Spain}

\author{Enric Canadell}
\affiliation{Institut de Ci\`encia de Materials de Barcelona (ICMAB-CSIC), Campus UAB, 08193 Bellaterra, Spain}

\begin{abstract}
  The origin of the charge density wave in TaTe$_4$ is discussed on the basis of a first-principles DFT analysis of the Fermi surface, electron-hole response function, phonon band structure of the average structure and structural optimization of the modulated phase. Analysis of the band structure and Fermi surface of the average structure clearly proves that despite the presence of TaTe$_4$ chains in the crystal structure, TaTe$_4$ is in fact a 3D material as far as the electronic structure near the Fermi level is concerned. A Fermi surface nesting mechanism is dismissed as the origin of the 2$a\times2a\times3c$ structural modulation. The optimized 2$a$\xone2$a$\xone3$c$ structure, which is found to be the more stable modulation in agreement with the experimental observations, can be obtained directly from a soft-phonon mode computed for the undistorted structure. Our results suggest that the driving force for the distortion is the maximization of Ta-Ta metal-metal bonding subject to inducing the minimum bonding decrease in the Te sublattice. 
\end{abstract}


\maketitle

\section{I. Introduction}\label{sec:intro}

The transition metal tetratellurides NbTe$_4$~\cite{Selte1964} and TaTe$_4$~\cite{Bjerkelund1984a,Bjerkelund1984} have recently been the object of considerable attention in the context of the search for materials with high magnetoresistance and the competition between charge density waves (CDWs) and superconductivity~\cite{Gao2017,Luo2017,Yang2018,Lima2019}. Transition metal tellurides often exhibit crystal structures and transport properties differing from those of the corresponding selenides or sulphides~\cite{Meerschaut1986}. The tellurium 5$p$ orbitals are considerably more diffuse than those of sulphur 3$p$ and selenium 4$p$ so that the valence bands in transition metal tellurides are wider than those of selenides or sulphides and may overlap substantially with the bottom part of the transition metal based $d$ bands. This leads to a non-negligible electron transfer from tellurium to the transition metal~\cite{Canadell1992} and as a consequence, the formal $d$ electron count for the transition metal atom is often not obvious. These electron transfers have strong implications for the structural and transport properties of many transition metal tellurides.   

\begin{figure}[htbp]
\centering
\includegraphics[scale=0.125]{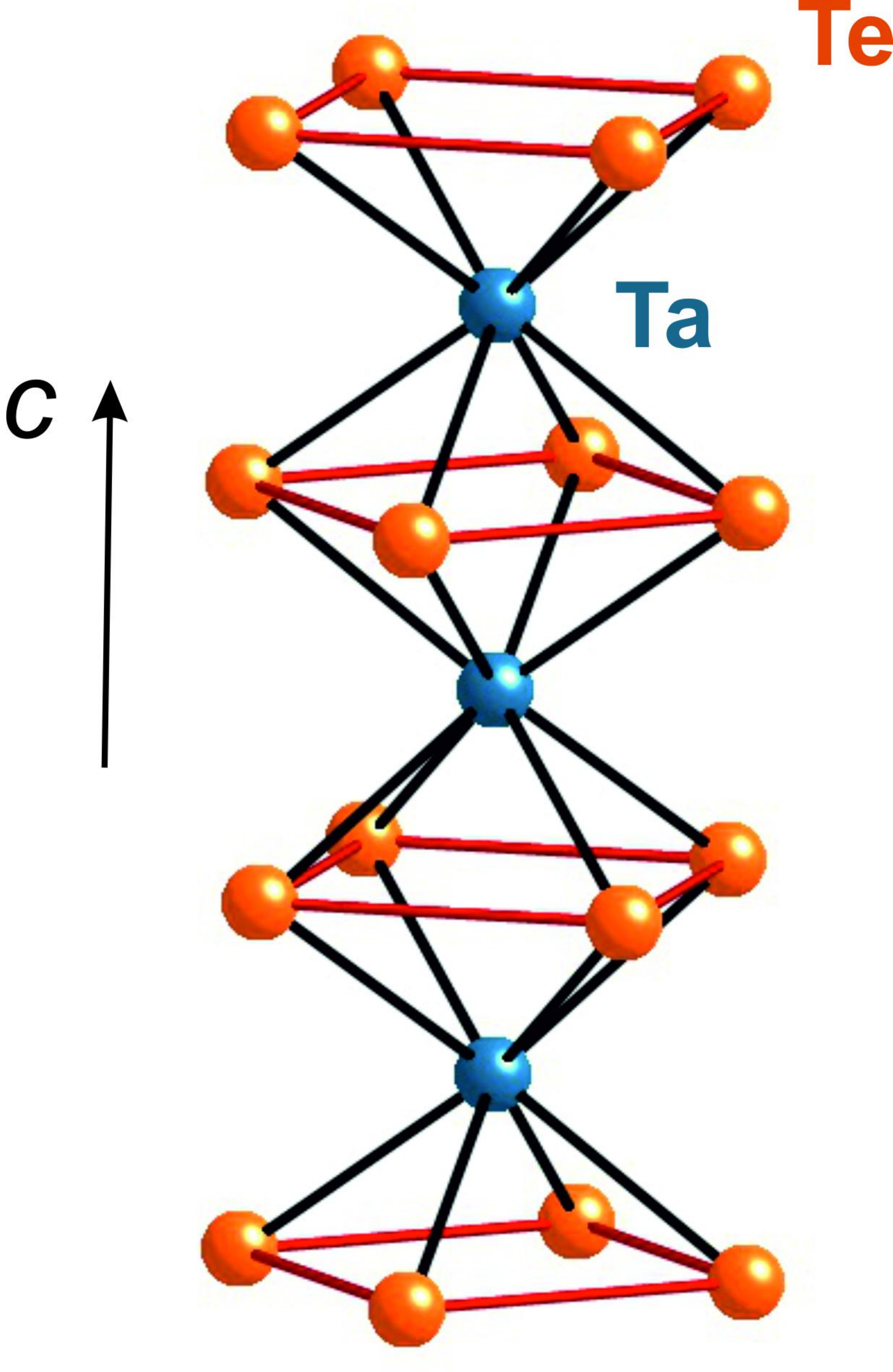}
\caption{Lateral view of a TaTe$_4$ chain where the Ta atoms are in a square antiprismatic coordination. }
\label{fig:tate4_structure_ac}
\end{figure}

Group V tetratelurides provide a clear example of such a situation. NbTe$_4$ and TaTe$_4$ are described as containing one-dimensional (1D) chains of transition metal atoms which are coordinated by the tellurium atoms in a square antiprismatic way (Fig.~\ref{fig:tate4_structure_ac}). VS$_4$ exhibits a related crystal structure built from VS$_4$ rectangular antiprismatic chains~\cite{Kozlova2015,Allman1964} and has the same number of valence electrons. VS$_4$ is a semiconductor as a result of the usual Peierls dimerization of a chain of transition metal atoms with a $d^1$ electron count (i.e. possessing a half-filled quasi-1D band). In contrast, NbTe$_4$ and TaTe$_4$ are metallic, their high temperature crystal structure contains uniform chains and exhibit structural modulations towards a 2$a\times2a\times3c$ commensurate phase. TaTe$_4$ has a transition to this commensurate phase at around 450 K \cite{Bronsema1987} while its isostructural counterpart, NbTe$_4$, crosses several incommensurate phases before locking into the final commensurate one at 50 K~\cite{Kusz1994,Smaalen1986}. Because of the presence of chains of transition metal atoms and the high-temperature metallic character, these CDWs have been attributed to a Peierls type mechanism as in other quasi-1D metals~\cite{Boswell1983,Whangbo1984,Tadaki1990}. However, in contrast with other quasi-1D chalcogenides exhibiting structural modulations like NbSe$_3$ or monoclinic-TaS$_3$~\cite{Meerschaut1986}, the resistivity hardly exhibits any change when the commensurate structural modulation sets in. Thus, the FS driven mechanism behind the CDW formation~\cite{Gruner1994} may not be at work here. More recently a different 4$a\times6c$ CDW has been shown to occur in the surface of the TaTe$_4$ crystals~\cite{Sun2020}. 

The origin of the CDW in these tetratellurides has been discussed for longtime. However, to the best of our knowledge there have not been first-principles studies discussing in detail the origin of these structural modulations until very recently. After completion of this work, a study was reported  concerning the Kohn anomaly in TaTe$_4$~\cite{Liu2021} although neither the distorted modulated structure nor the correlation between the structural and electronic properties were considered in detail. A Density Functional Theory (DFT) study by Bullet~\cite{Bullet1984} noted the unusual strength of inter-chain interactions and an early tight-binding study~\cite{Whangbo1984} proposed that FS nesting was indeed at work although the reported FS does not provide compelling evidence for such claim. A more recent DFT study provided data concerning the electronic structure of TaTe$_4$ but the origin of the bulk CDW was not considered~\cite{Sun2020}. Angular resolved photoemission (ARPES) spectra of TaTe$_4$ exhibits a combination of 1D and 3D Fermi surface components~\cite{Zwick1999}: the experiment does not clarify though if there are two decoupled Fermi surface components, nor if the Fermi surface nesting plays a role in the structural modulation although a recent theoretical study suggests that electron-phonon is the main source of the modulation~\cite{Liu2021}. Note that the electronic structure of the room temperature modulated structure has not been considered in any of these works. Because of the renewed interest on these solids~\cite{Gao2017,Luo2017,Yang2018,Lima2019,Sun2020}, we decided to carry out a first-principles DFT study for the average and modulated structures of TaTe$_4$ by means of an analysis of the FS, electron-hole response function and phonon spectra of the non-modulated structure and structural optimizations of the CDW phase. We focus in TaTe$_4$ although very similar arguments should apply for NbTe$_4$.

\section{II. Computational details}\label{sec:compdet}

The DFT calculations~\cite{HohKoh1964,KohSha1965} were carried out using a numerical atomic orbitals DFT approach implemented in the \textsc{Siesta} code~\cite{SolArt2002,ArtAng2008,Garcia2020}. The Perdew-Burke-Ernzerhof (PBE) functional was used to account for the exchange-correlation energy~\cite{PBE96}. The core electrons have been replaced by norm-conserving scalar relativistic pseudopotentials~\cite{tro91} factorized in the Kleinman-Bylander form~\cite{klby82}. We have used a split-valence double-$\zeta $ basis set including polarization functions~\cite{arsan99}. In all calculations, we use a cutoff of 800 Ry for the real space integrals, and a tolerance of $10^{-4}$ and $10^{-3}$ eV on the density matrix and the total energy, respectively, for the convergence of the SCF cycle. To sample the Brillouin zone (BZ) for the electronic states, a Monkhorst-Pack~\cite{MonPac76} $k$-points grid of 32$\times$32$\times$30 was used for the non-distorted unit cell and scaled accordingly to the supercell calculations. The Lindhard electron-hole response function,

\begin{equation}\label{eq:chi}
\chi(q)=-\sum_{i,j}\sum_{k}\frac{f_F(\epsilon_i({k}))-f_F(\epsilon_j({k}+{q}))}{\epsilon_i({k})-\epsilon_j({k}+{q})},
\end{equation}

\noindent
where \textit {f}$_F$ is the Fermi-Dirac distribution function, was obtained from the computed DFT band eigenvalues $\epsilon_i({k})$. The integral over {\it k}-points of the BZ was approximated by a direct summation over a dense, regular grid of points. As the Lindhard function is more sensitive to the accuracy of the BZ integration than the total energy, the {\it k}-points grid used for its calculation must be more dense than in the standard self-consistent determination of the charge density and Kohn-Sham energy. The calculations are done using the eigenvalues obtained in the DFT calculation for the coarser grid, and interpolating their values in the denser grid, using a post-processing utility available within the \textsc{Siesta} package. In this work, the BZ was sampled using a grid of (200$\times$200$\times$200) {\it k}-points, for the calculation of the Lindhard response function. Phonon calculations were carried out using the finite difference method available in the SIESTA code. In the case of the phonon band structure, a {\it k}-point grid of 10$\times$6$\times$10 per minimum unit cell in a 5$\times$3$\times$5 supercell was used. The Fermi-Dirac smearing was set to $5\times10^{-3}$ eV.

\section{III. Crystal structure and electron counting}\label{sec:crystal_structure}

\subsection{A. Crystal Structure}

\begin{figure}[htbp]
\centering
\protect\includegraphics[scale=0.125]{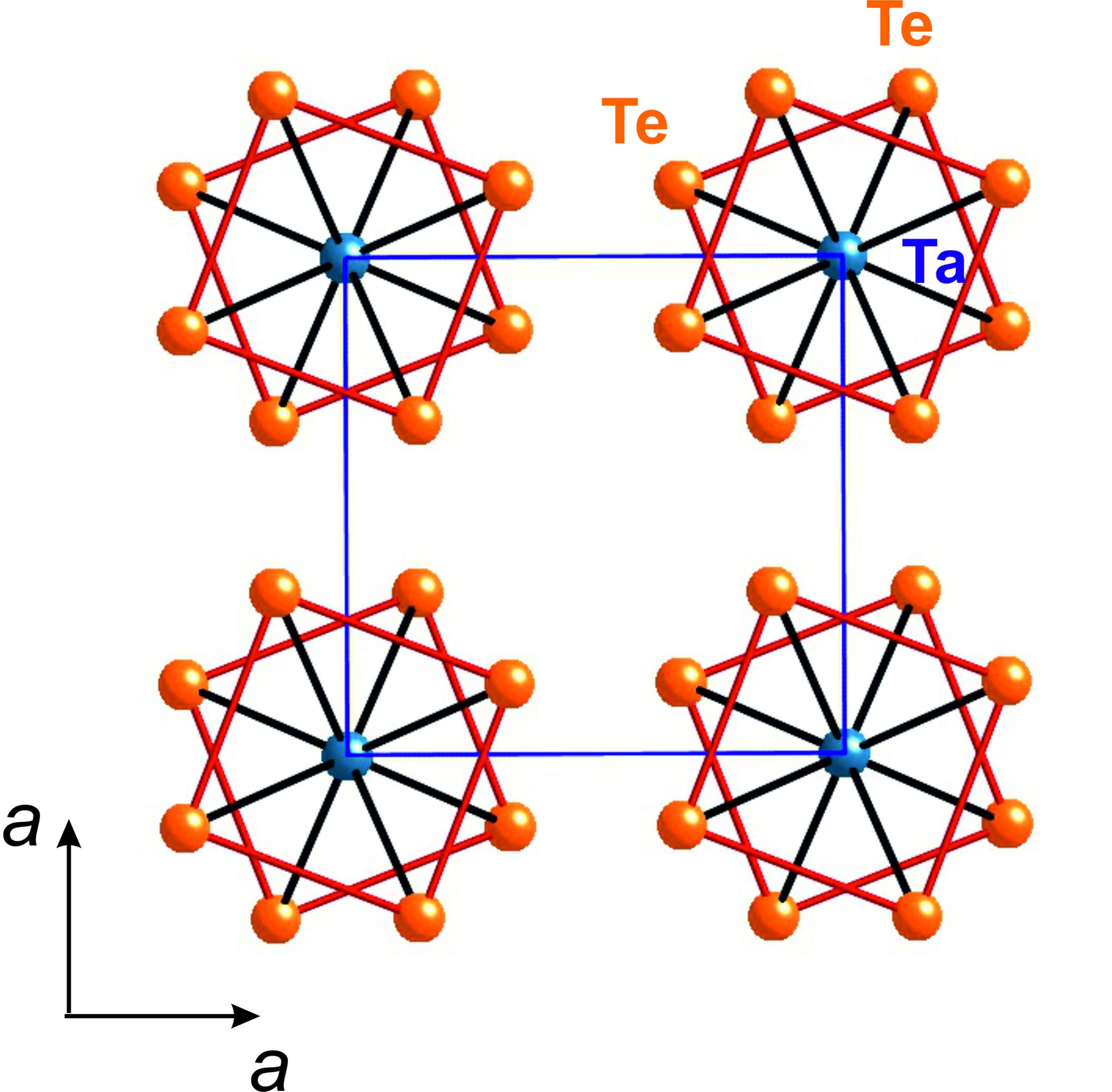}
\caption{Top view of the tetragonal structure of TaTe$_4$. Te atoms are represented in orange and Ta in blue. Red lines represent Te-Te contacts in the shared faces of the square antiprisms.}
\label{fig:tate4_structure_aa}
\end{figure}

TaTe$_4$ crystallizes in a tetragonal structure with Ta atoms coordinated by 8 Te atoms in a chain like fashion. The structure of the 2$a\times2a\times3c$ commensurately modulated phase was solved by Bronsema et al.~\cite{Bronsema1987}. In the average structure determined by these authors the Ta atoms are located in the center of a square
antiprism of Te atoms, where the two Te$_4$ square units are rotated 45$^{\circ}$ with respect to each other as shown in Fig. \ref{fig:tate4_structure_ac}. The crystal
structure of TaTe$_4$ is a tetragonal array of chains of this type (Fig. \ref{fig:tate4_structure_aa}). The space group symmetry of the average structure has been the matter of some discussion~\cite{Bjerkelund1984,Boswell1983,Eaglesham1985} but it was finally determined by Bronsema et al.~\cite{Bronsema1987} to be $P4/mcc$ with the lattice parameters $a$= 6.5154 \AA~and $c$= 6.8118 \AA. Their corresponding Wyckoff positions are 2a for Ta atoms situated in (0, 0, 1/4) and (0, 0, 3/4) in fractional units, respectively 8m for Te atoms situated in ($x$, $y$, 0) and its equivalent positions, where $x$ = 0.1438 and $y$ = 0.3274. We carried out a relaxation of the internal coordinates while forcing the space group symmetry giving us $x$ = 0.1501 and $y$ = 0.3296, values which are in good agreement with the experimental ones. The space group of the 2$a\times2a\times3c$ commensurately modulated phase was found to be $P4/ncc$~\cite{Bronsema1987}. Our DFT structural optimization is in excellent agreement with this work (see Sect. V). Let us note that weak superlattice reflections at (1/2, 0, 1/3), (0, 1/2, 1/3), and very weak ones at (0, 0, 1/3) were also observed~\cite{Boswell1983,Bronsema1987} although the last ones are most likely second harmonics satellites.   

\subsection{B. Electron counting}\label{sec:el_count}

\begin{figure}[!hptb]
    \centering
    \includegraphics[width=0.40\textwidth]{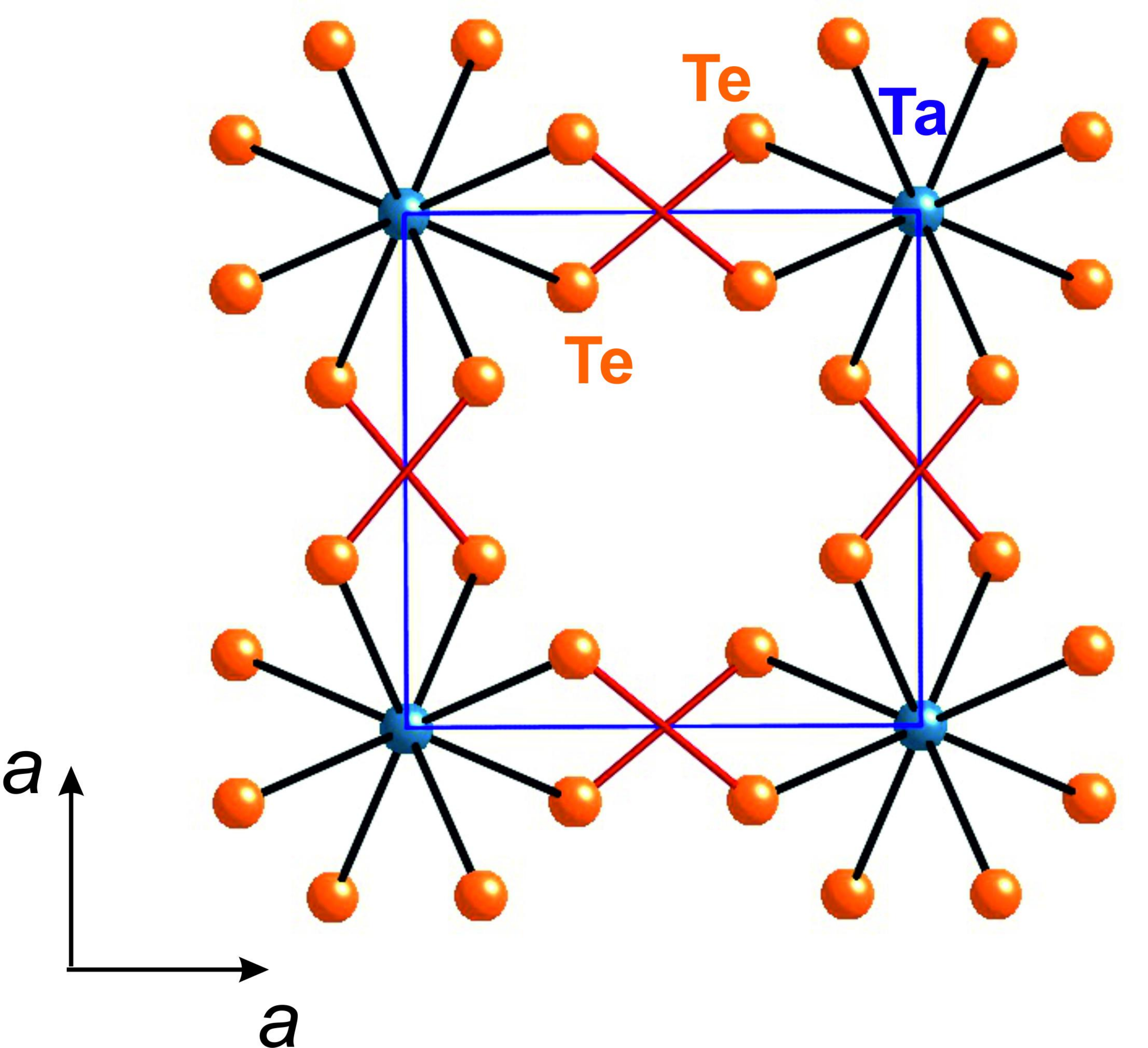}
    \caption{Top view of the tetragonal structure of TaTe$_4$. Te atoms are represented in orange and Ta in blue. Red lines represent the shorter Te-Te bonds of the structure (2.93 \AA~in the average crystal structure~ \cite{Bronsema1987}).}
    \label{fig:tate4_3D}
\end{figure}

For a better understanding of the structural modulation, we need to determine how many electrons fill the Ta-based bands in the high temperature structure. The first thing to do is to look at the Te$-$Te distances. Those within a Te$_4$ square unit (3.29 \AA) are shorter than the sum of the van der Waals radii of Te (4.0 \AA), but too long to be considered as a real Te-Te bond (2.7 - 2.9 \AA). At a careful inspection of the structure we can observe that the shorter Te-Te contacts are not those in the Te$_4$ square units but those connecting the squares of neighbouring chains (the red bonds in Fig. \ref{fig:tate4_3D}, 2.93 \AA), which are of the same order as many Te-Te single bonds. Under these considerations the structure should be viewed as a 3D lattice of Ta-Te and Te-Te bonds as shown in Fig. \ref{fig:tate4_3D}. This is in contrast with the more usual description of this structure as a tetragonal array of TaTe$_4$ chains (see Fig.~\ref{fig:tate4_structure_aa})~\cite{Selte1964,Boswell1983,Meerschaut1986}. It is also in contrast with the VS$_4$ structure~\cite{Kozlova2015,Allman1964} where there are no short S-S contacts between the chains but within the chains (i.e. the short sides of the rectangular S$_4$ groups separating the successive V atoms, see Fig. S2 in the Supporting Information (SI)).

Because of the occurrence of Te-Te bonds the tellurium atoms should be considered as (Te$_2$)$^{2-}$, which means that only one electron is left to fill  the Ta-based bands (i.e. Ta atoms are formally $d^{1}$). Thus only the bottom Ta 5$d$ based bands may be partially filled. The $d$ orbitals of a transition metal in this coordination environment are such that the lowest energy orbital is a $d_{z^{2}}$ orbital and slightly higher in energy there are the two $d_{xy}$ and $d_{x^{2}-y^{2}}$ orbitals. These two orbitals will be very weakly dispersive along the Ta chains direction whereas the $d_{z^{2}}$ orbital creates strong interactions and will thus lead to a wide band. Consequently, the bottom Ta-based band will be a dispersive 5$d_{z^{2}}$ band. Since there is just one electron to fill this band, the bottom part of the Ta based bands will be most likely a half-filled Ta 5$d_{z^{2}}$ band.

On the basis of this formal electron count we can predict that if the solid behaves as a 1D system around the Fermi level, it should be unstable to some kind of dimerization that would open a gap at the Fermi level, as it is the case for VS$_4$~\cite{Kozlova2015}. However, this is in conflict with the experimental observations. To begin with the modulated structure is metallic. Moreover the observed superlattice spots all have a 1/3 component along the $c$-direction (the direction of the Ta chains). Thus, if the system can be treated as electronically 1D, the Ta band should not be half-filled but 1/3rd or 2/3rd filled. 

\begin{figure*}[t]
    \centering
    \includegraphics[width=0.95\textwidth]{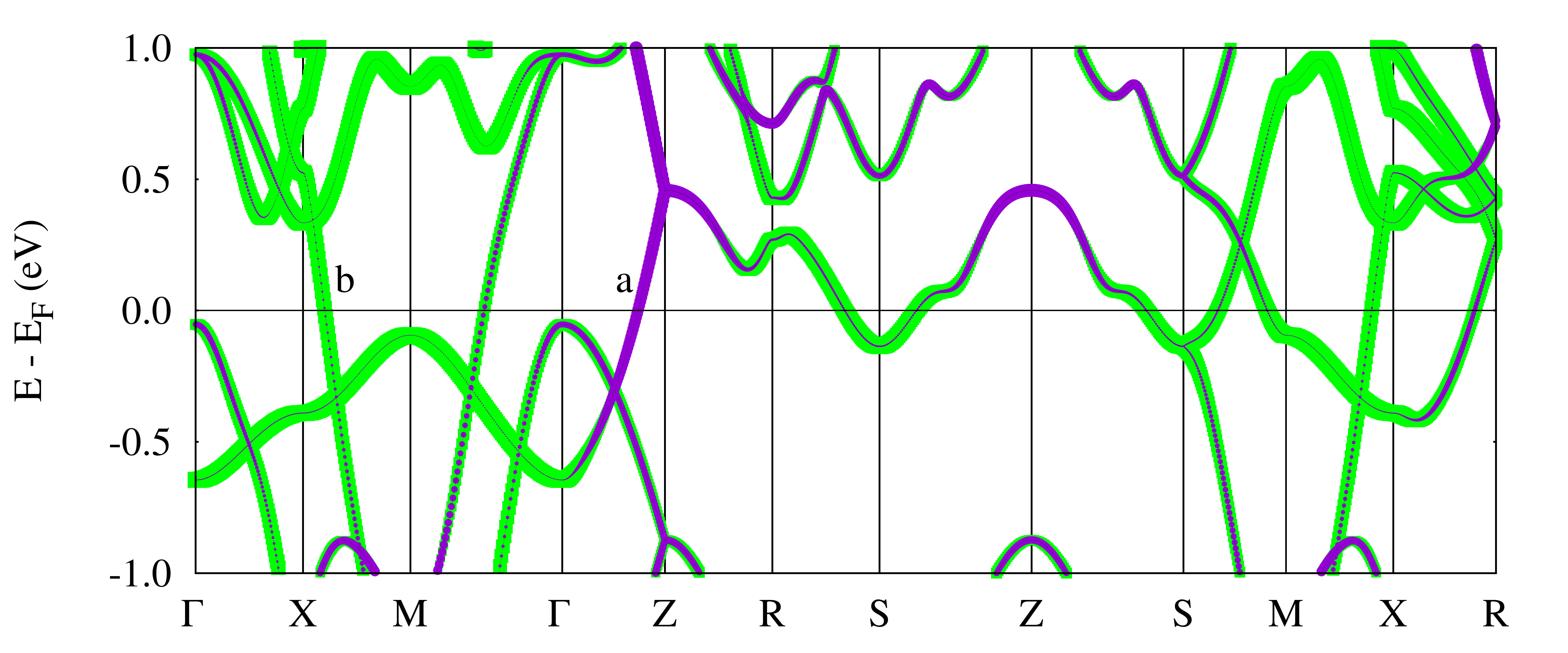}
    \caption{Fatband structure of the TaTe$_4$ average structure (see Fig. \ref{fig:tate4_fs}d for the labeling of the special points). The size of the green and purple dots are proportional to the Te 5$p$ and Ta 5$d_{z^2}$ character, respectively.}
    \label{fig:tate4_fat}
\end{figure*}

The latter possibility can be easily understood: it simply means that there is some electron transfer from the Te valence band towards the Ta 5$d_{z^{2}}$ band. This is indeed likely because of the short Te-Te contacts between the Te$_2$ units which may raise the top of the Te valence band. Thus, the top of the valence band may end up being higher than the Fermi level, transferring electrons to the Ta 5$d_{z^{2}}$ band which can thus become 2/3rd filled. 


In contrast, the first possibility may appear at the first sight to be unlikely because according to the occurrence of (Te$_2$)$^{2-}$ units, the Te valence band is already full and can not accept electrons from Ta. To acquire electrons from the Ta atoms part of the Te-Te bonds should be broken, generating two Te$^{2-}$, which is not the case according to the crystal structure~\cite{Bronsema1987}. However, let us remind that although the Te-Te contacts of the Te$_4$ squares are longer than those between the squares, they are considerably shorter than the sum of the van der Waals radii. Taking into account both types of Te-Te contacts, one can see that one ($aa$) plane of Te atoms is made of a series of strongly interacting Te-Te bonds (see Fig. S1 in SI) so that there must be very strong interactions between the Te-Te bonds along the layer. In that way the empty band originating from the antibonding $\sigma$* Te-Te levels will broaden and the bottom part could overlap with the filled 5$d$ levels. The result would be a Ta to Te electron transfer leading to a less than half-filled Ta 5$d_{z^{2}}$ band.  

The previous considerations raise some serious questions impinging directly on the CDW mechanism: do the Te-Te short inter-chain contacts confer the electronic structure around the Fermi level with a 2D or 3D character? In that case the probability to have good Fermi surface nesting is unlikely and the CDW would not be Fermi surface nesting driven. In addition, it is doubtful that the electronic transfer leads to a commensurate filling of the 5$d_{z^{2}}$ band. This objection can be somewhat dismissed by noting that the isostructural and isoelectronic NbTe$_4$ undergoes a series of incommensurate modulations along the $c$-direction before becoming commensurate at low temperature. Thus, maybe in TaTe$_4$ (but not in NbTe$_4$) the electron transfer is very near the commensurability. Note also that whatever it is the sense of the electron transfer, even if the Ta-based bands lead to a Fermi surface nesting driven CDW and thus to the opening of a gap at the Fermi level, the Te-based bands may not be necessarily affected so that the metallic character of the bands can be kept after the CDW sets in the material, as it is in fact observed. Thus, it is not at all clear from simple electron counting arguments that the modulation exhibited by TaTe$_4$ originates from a Fermi surface nesting phenomena, as in other metallic low-dimensional materials, or from a phonon driven instability, as for instance in 2$H$-NbSe$_2$~\cite{Johannes2006}.

\section{IV. Electronic structure of the non-modulated structure}\label{sec:elec_struct_ave}
\subsection{A. Bands crossing the Fermi level}\label{sec:BS_ave}

\begin{figure*}[!hptb]
    \centering
    \includegraphics[width=0.75\textwidth]{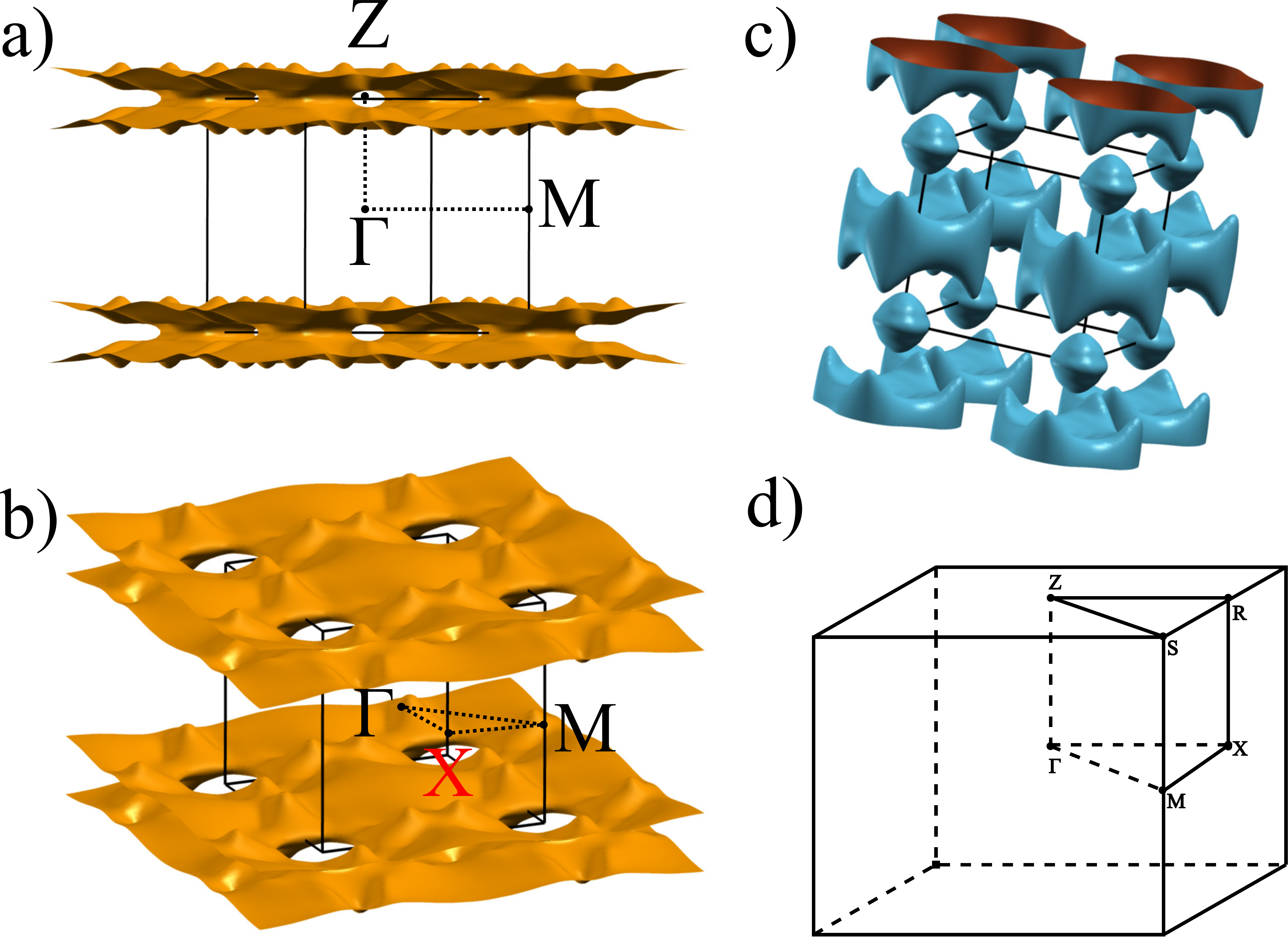}
    \caption{Fermi surface of the TaTe$_4$ average structure. (a) and (b) are two views of the component mostly originating in the Ta 5$d_z^{2}$ orbitals. (c) Component originating in the Te 5$p$ orbitals. d) Brillouin zone (BZ) of a tetragonal lattice. The high symmetry points have the following coordinates: $\Gamma = (0,0,0), X = (0.5,0,0), M = (0.5,0.5,0), Z = (0,0,0.5), R = (0.5,0,0.5)$ and $S = (0.5,0.5,0.5)$ in units of the reciprocal lattice vectors.}
    \label{fig:tate4_fs}
\end{figure*}

The calculated band structure for the non-modulated structure of TaTe$_4$ is shown in Fig.~\ref{fig:tate4_fat}. The weighted band dispersion (fatband) analysis reveals that there are essentially two types of bands crossing the Fermi level. The band crossing from $\Gamma$ to Z (noted $a$ in the figure) originates from Ta 5$d_{z^2}$ orbitals and is actually a folded band because there are two Ta atoms in the unit cell related by a screw axis along the $c$ direction. Since the lower band component is only partially filled it appears that the formal electron count for Ta is lower than $d^{1}$.
Note also that the Te content of this pair of bands is quite sizeable and in several lines of Fig.~\ref{fig:tate4_fat} even seems to dominate. However this is partially due to the 1:4 stoichiometry which increases the total weight of the Te orbitals. What is however quite clear is that whereas these Ta 5$d_{z^2}$ bands are strongly dispersive along the $c$ direction they are also engaged in inter-chain interactions (see for instance the $\Gamma$-M-X-$\Gamma$ lines in Fig.~\ref{fig:tate4_fat}) because of the mixed Ta and Te character and the short inter-chain Te-Te contacts highlighted in Fig.~\ref{fig:tate4_3D}. This band should thus have a strong memory of the 1D type interactions associated with the Ta 5$d_{z^2}$-Ta 5$d_{z^2}$ interactions along the chain while exhibiting a non-negligible warping due to the tellurium content and the inter-chain interactions.

There is a second band crossing the Fermi level (noted $b$ in the figure) which is mostly based on Te 5$p$ orbitals and exhibits a large dispersion on both directions, perpendicular (see $\Gamma$-M and M-X in Fig.~\ref{fig:tate4_fat}) and parallel (see M-S in Fig.~\ref{fig:tate4_fat}) to the TaTe$_4$ chains. This mostly Te-based band should thus have a 3D character. In fact, the two bands overlap and interact quite strongly, thus interchanging character along different parts of the Brillouin zone (see for instance the Z-S line in Fig.~\ref{fig:tate4_fat}). Thus, one should expect a relatively complex FS resulting from the interaction between the Ta 5$d_{z^2}$ and Te based bands for non-modulated TaTe$_4$.

\subsection{B. Fermi surface and electron-hole response}\label{sec:FS_ave}

The FS of TaTe$_4$ is shown in Fig. \ref{fig:tate4_fs}. It is mostly made of two different contributions originating from the two types of bands ($a$ and $b$) mentioned above. As expected, one of the components is relatively flat (Figs.~\ref{fig:tate4_fs}a and b) although bearing a very substantial degree of warping and even, holes. In fact, the two sheets making this portion of the FS touch at many points of the Brillouin zone thus making a complex slab with holes and closed empty regions inside. Coming to the second component of the FS, we notice that it has a three-dimensional topology since the band generating this FS component crosses the Fermi level in several directions like $\Gamma$-M, M-X, but also along M-S where the band mixes with the Ta based band. In fact, because of the degeneracy of the bands in the outer plane of the Brillouin zone (see the traject Z-R-S-Z in Fig.~\ref{fig:tate4_fat}) there is a third Te 5 $p$ based band crossing the Fermi level leading to the closed almost spherical pockets around the corners of the Brillouin zone (point S) in Fig.~\ref{fig:tate4_fs}c.

\begin{figure*}[!hptb]
    \centering
    \includegraphics[width=0.75\textwidth]{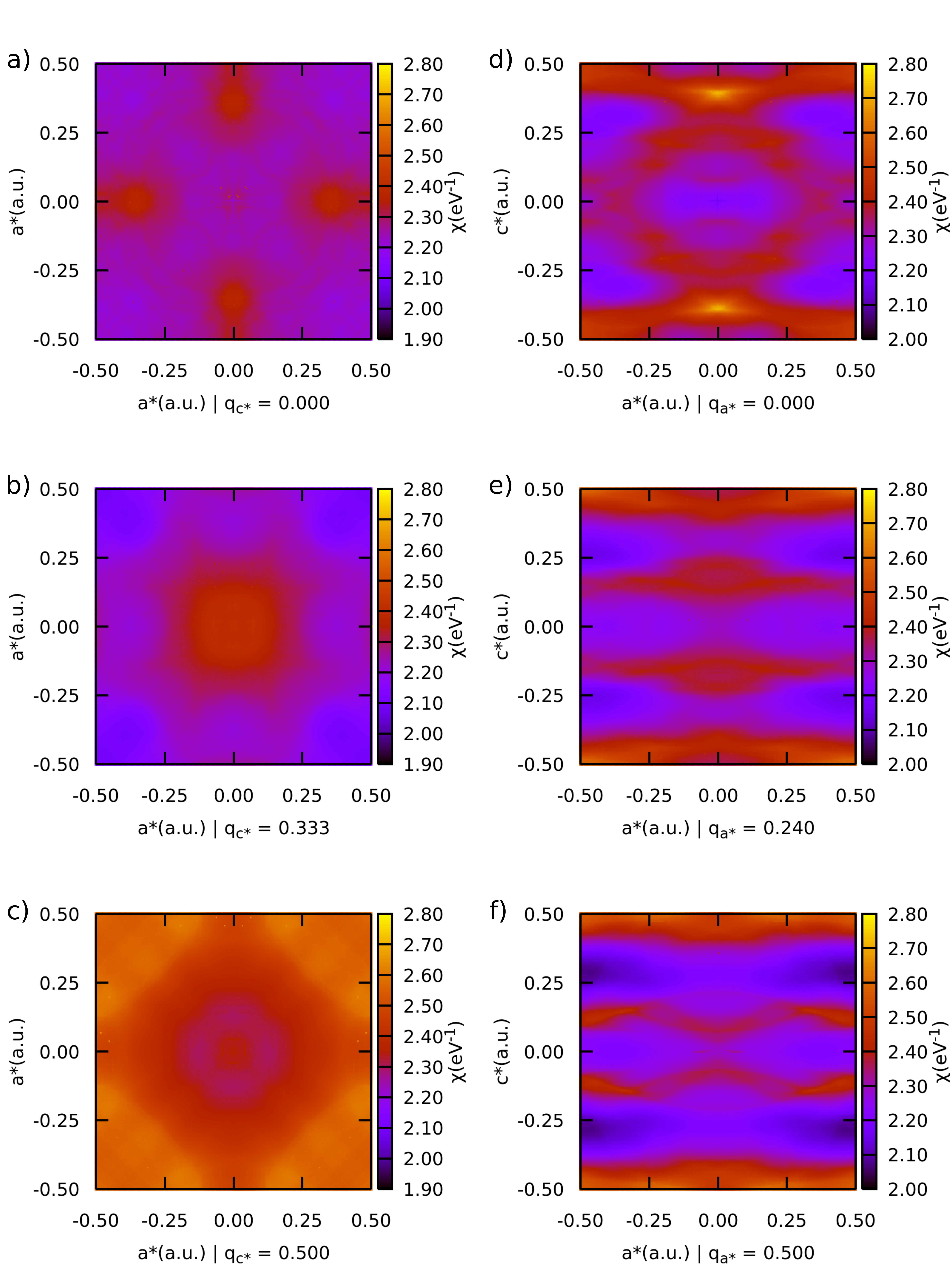}
        \caption{2D mapping of the calculated Lindhard function of TaTe$_4$ in the a*-a* plane for different $q_{c^*}$ components (a-c), and in the $a$*-$c$* plane for different secondary $q_{a^*}$ components (d-f). The bars represent the Lindhard function value in arbitrary units.}
    \label{fig:tate4_lindhard}
\end{figure*}

The warped FS component of Fig.~\ref{fig:tate4_fs}a, which contains substantial flat portions, could lead to non-negligible nesting and thus be at the origin of the CDW. This FS component, originating from band $a$ in Fig.~\ref{fig:tate4_fat}, is associated with less than 2 electrons per unit cell (which is the filling associated with Ta atoms with an electron counting of $d^{1}$). This result confirms that, in contrast with the usual situation in many transition metal tellurides, there has been a formal electron transfer from the Ta to the Te sublattices, i.e. the formal electron counting for Ta is $d^{1-\delta}$ with $\delta\sim$ 0.2, in between those of a one-third filled ($\delta$= 2/3) and half-filled ($\delta$= 0). It is because of this transfer that the "double pillow" pockets of Fig.~\ref{fig:tate4_fs}c are created. As mentioned above, this reverse electron transfer means that the tellurium sublattice can not be considered as a series of isolated (i.e. not $directly$ interacting) Te-Te bonds (i.e. (Te$_2$)$^{2-}$). A series of Te-Te bonds is only compatible with $\delta$= 0. The filling of additional Te based levels implicates that some of the antibonding and thus empty Te-Te levels have been stabilized so as to occur in the region of the Fermi level and are thus filled. This can occur when the Te-Te interaction is considerably delocalized through the tellurium sublattice because then, the Te-Te empty antibonding levels interact and spread into wide bands because of the direct Te-Te inter-bond interactions, the bottom part of such bands (which have some bonding character between the Te-Te units, i.e. the dotted lines in Fig. S1 of SI) overlaps with filled transition metal levels leading to a reverse electron transfer from the transition metal to tellurium (and to some bonding between Te-Te bonds). A consequence of these electron transfers is that the formal electron counting for the tellurium atoms and fragments of transition metal tellurides is sometimes not quite obvious~\cite{Papoian2000,Patschke2002}. As a consequence of these transfers the anion sublattice of transition metal tellurides often contains tellurium fragments larger than dimers and even large polianionic units~\cite{Canadell1992b,Patschke2002}. The present results clearly confirm that the tellurium layers of TaTe$_4$ should be regarded as a strongly interconnected lattice of Te$_2$ units, as sketched in Fig. S1 (SI).

In order to clarify if the structural modulation originates in a Fermi surface nesting phenomena we have performed Lindhard function calculations across several planes in both ($a^{*}a^{*}$) and ($a^{*}c^{*}$). According to the experimental findings, if the CDW modulation originates from FS nesting we should find sharp maxima in the Lindhard function calculations for $1/2a^{*}$ and $1/3c^{*}$ components, i.e. in the corners of Fig. \ref{fig:tate4_lindhard}b. However, this is not the case. For the ($a^{*}a^{*}$) planes when we vary the $q_{c^{*}}$ component, the maximum in the Lindhard function occurs for $q_{c^{*}} = 1/2$ (Fig. \ref{fig:tate4_lindhard}c) and not $1/3$. Even so, the occurring maximum is not a typical logarithmic dispersion for a Fermi surface nesting driven instability, but it covers a rather broad region of suitable nesting vectors for $q_{a^{*}}$ in the $[0.1-0.5]$ interval. It could be argued that the 1/2 components of the modulation along the inter-chain directions originate from Coulomb coupling between intra-chain modulations. However, in the case where we calculate the Lindhard function for ($a^{*}c^{*}$) planes and vary the second $q_{a^{*}}$ component (Figs.~\ref{fig:tate4_lindhard}d-f), we find that the maximum response is given by $q = (0,0,0.39)$ (see Fig.~\ref{fig:tate4_lindhard}d). This wave vector does not correspond to the 2k$_F$ value of a folded d$_{z^2}$ band for Ta $\sim d^{0.8}$, i.e. is not related to the Ta 5$d_{z^2}$ based FS component of Fig.~\ref{fig:tate4_fs}a. In fact, this wave vector is associated with an inter-band nesting and would lead to an incommensurate modulation along the $c$ direction which is not a proper nesting vector to explain the observed modulation. We thus may dismiss the possibility of a FS driven instability leading to the structural modulation manifested in TaTe$_4$.


\section{V. Structural modulation}\label{sec:modulation}
Assuming no prior knowledge of the modulated structure of this material, the clear-cut answer of phase stability/ordering, whichever they might be, is provided by the energy gain of each phase with respect to the undistorted structure. In this section we examine this point by calculating the phonon dispersion and carrying out actual structural optimizations for different modulated structures.

\subsection{A. Phonon band structure}\label{sec:phonon}

\begin{figure*}[htbp]
\centering
\includegraphics[width=0.80\textwidth]{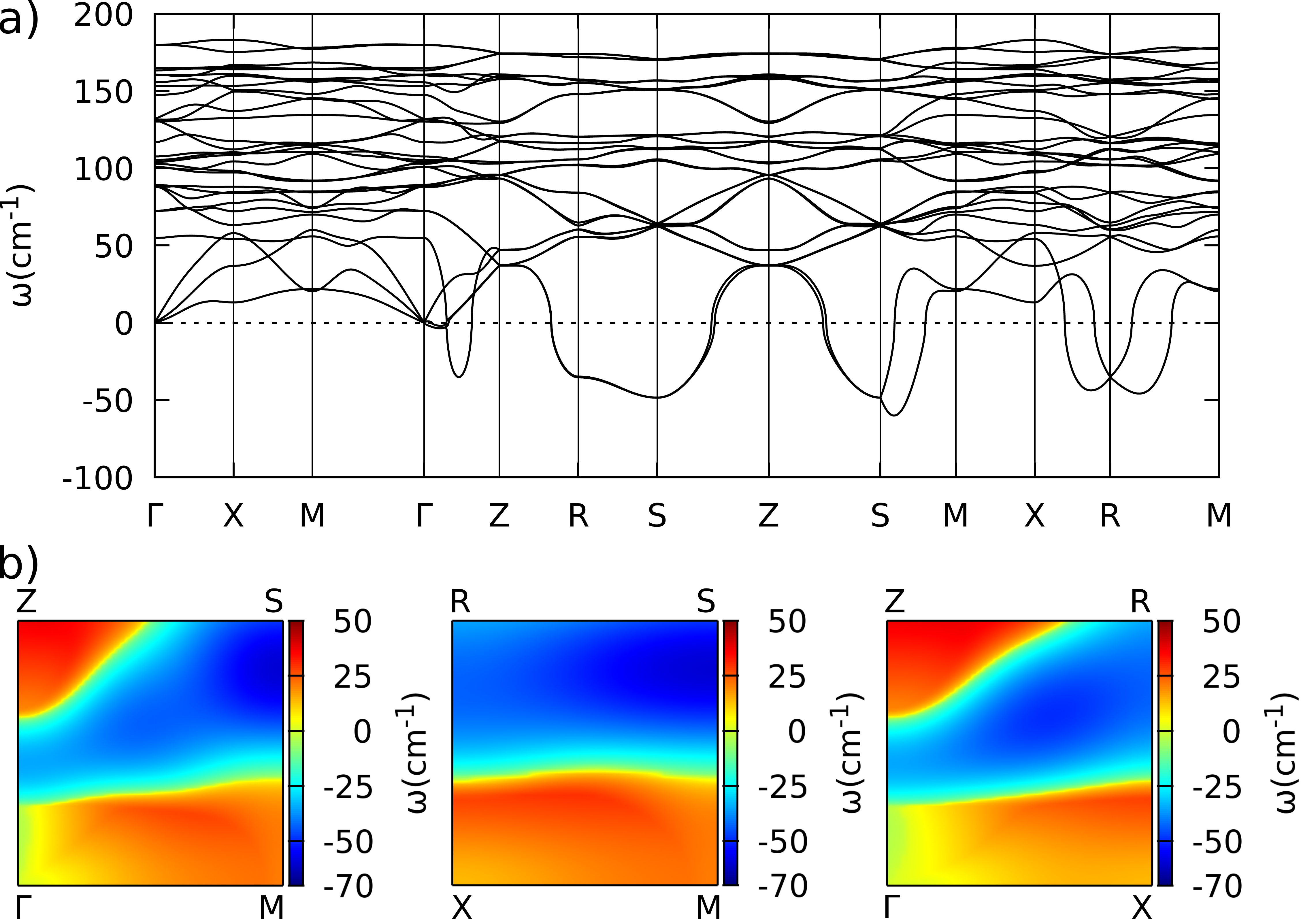}
\caption{Phonon band structure along high-symmetry lines (a). Landscape of soft-phonon frequencies along high-symmetry planes in reciprocal space. The lowest energy phonon mode for each k-point is represented (b). }
\label{fig:phonons}
\end{figure*}

The vibrational modes in the undistorted phase can give information to understand the origin of the possible structural modulations. The calculated phonon band structure is shown in figure \ref{fig:phonons}.
%
Several imaginary phonon modes can be identified, with minima not too far from (0,0,$\frac{1}{4}$), ($\frac{1}{2}$,$\frac{1}{2}$,$\frac{1}{3}$), (0,$\frac{1}{2}$,$\frac{1}{3}$). Indeed, this complex phonon band structure is a  consequence of a landscape of lattice instabilities with a valley of modulations along the $c^*$ axis which is between $\frac{1}{4}$ and $\frac{1}{3}$, indicating that multiple possible phases might coexist.
Guided by the soft-phonon minima, we have performed internal coordinates relaxations for each of the possible phases, and the lowest energy state we obtained was the 2$a$\xone2$a$\xone3$c$ phase, in agreement with experimental observations. All the other stable phases were higher in energy than the 2$a$\xone2$a$\xone3$c$ phase as presented in Table \ref{table:tate4_de}. An optimization for the 1$a$\xone1$a$\xone3$c$ phase led to the average structure. We also considered a $\sqrt{2}a$\xone$\sqrt{2}a$\xone3$c$ phase. The energy gain for this structure was found to be half that of the 2$a$\xone2$a$\xone3$c$ phase; this is consistent with the fact that this modulation is only found for temperatures higher than 450 K.

\begin{table}[!hptb]
\centering
\caption{Energy gain of different commensurate superstructures
with respect to the TaTe$_4$ average structure.\\}
\begin{tabular}{|c|c|}
\hline
Periodicity & Energy gain(meV/f.u.) \\ \hline
1$a$\xone1$a$\xone4$c$    &  -13                  \\ \hline
$\sqrt{2}a$\xone$\sqrt{2}a$\xone3$c$    &  -28                  \\ \hline
2$a$\xone2$a$\xone4$c$    &  -31                  \\ \hline
2$a$\xone1$a$\xone3$c$    &  -38                  \\ \hline
2$a$\xone2$a$\xone3$c$    &  -57                  \\ \hline
\end{tabular}
\label{table:tate4_de}
\end{table}

\subsection{B. Modulated structure}\label{sec:DFT_optim_CDW}

The charge density wave structure and lattice symmetry have been determined experimentally by Bronsema and coworkers~\cite{Bronsema1987}. It was later restated by Corbett et al.~\cite{Corbett1988} that the modulated structure symmetry is indeed P4/ncc. We present in Table S-I of the SI a comparison between the experimental structure and our calculated structure of the lowest energy state. The calculated structure is in excellent agreement with the experimental one.
\begin{table}[!hptb]
\centering
\caption{Ta-Ta contacts in the optimized average structure and in the two types of chains in the optimized modulated structure. Note that the distortion is allowed within the experimental lattice parameters. Refer to Fig. \ref{fig:tate4_distance_labels} for labels.\\}
\begin{tabular}{|c|c|c|c|l|}
\hline
d(\AA)                  & Average                & Type A & Type B & \#  \\ \hline
\multirow{3}{*}{Ta-Ta} & \multirow{3}{*}{3.406} & 3.156  & 3.952  & (a) \\ \cline{3-5} 
                       &                        & 3.101  & 3.133  & (b) \\ \cline{3-5} 
                       &                        & 3.960  & 3.133  & (c) \\ \hline
\end{tabular}
\label{table:TaTa}
\end{table}
\begin{table}[!hptb]
\centering
\caption{Te-Te contacts in the optimized average structure and in the two types of chains in the optimized modulated structure. Both the interchain and the intrachain distances are included. A particular case represents chain type B where the square symmetry is broken and we have two types of distances within the rectangle. Moreover the planes containing these two consecutive rectangles are rotated by 90$^{\circ}$ with respect to each other (Thus the order of writing in the table). Note that the distortion is allowed within the experimental lattice parameters. Refer to Fig. \ref{fig:tate4_distance_labels} for labels.\\}
\begin{tabular}{|c|c|c|c|c|c|c|}
\hline
d(\AA)                  & Average                & Interchain & Average                & Square[A]  & Square[B]     & \#    \\ \hline
\multirow{3}{*}{Te-Te} & \multirow{3}{*}{2.959} & 2.974      & \multirow{3}{*}{3.337} & 3.143       & 3.389 - 3.471 & S$_a$ \\ \cline{3-3} \cline{5-7} 
                       &                        & 2.965      &                        & 3.414       & 3.471 - 3.389 & S$_b$ \\ \cline{3-3} \cline{5-7} 
                       &                        & 2.929      &                        & 3.447       & 3.146         & S$_c$ \\ \hline
\end{tabular}
\label{table:TeTe}
\end{table}
\begin{table}[!hptb]
\centering
\caption{Ta-Te contacts in the optimized average structure and in the two types of chains in the optimized modulated structure. Note that the distortion is allowed within the experimental lattice parameters. Refer to Fig. \ref{fig:tate4_distance_labels} for labels.\\}
\begin{tabular}{|c|c|c|c|l|}
\hline
d(\AA)  & Average                & Type A & Type B & \# \\ \hline
\multirow{6}{*}{Ta-Te} & \multirow{6}{*}{2.910} & 2.845  & 2.969  & 1  \\ \cline{3-5} 
                       &                        & 2.982  & 2.843  & 2  \\ \cline{3-5} 
                       &                        & 2.971  & 2.945  & 3  \\ \cline{3-5} 
                       &                        & 2.836  & 2.926  & 4  \\ \cline{3-5} 
                       &                        & 2.935  & 2.839  & 5  \\ \cline{3-5} 
                       &                        & 2.934  & 2.982  & 6  \\ \hline
\end{tabular}
\label{table:TaTe}
\end{table}

\begin{figure*}[!hptb]
    \centering
    \includegraphics[width=0.425\textwidth]{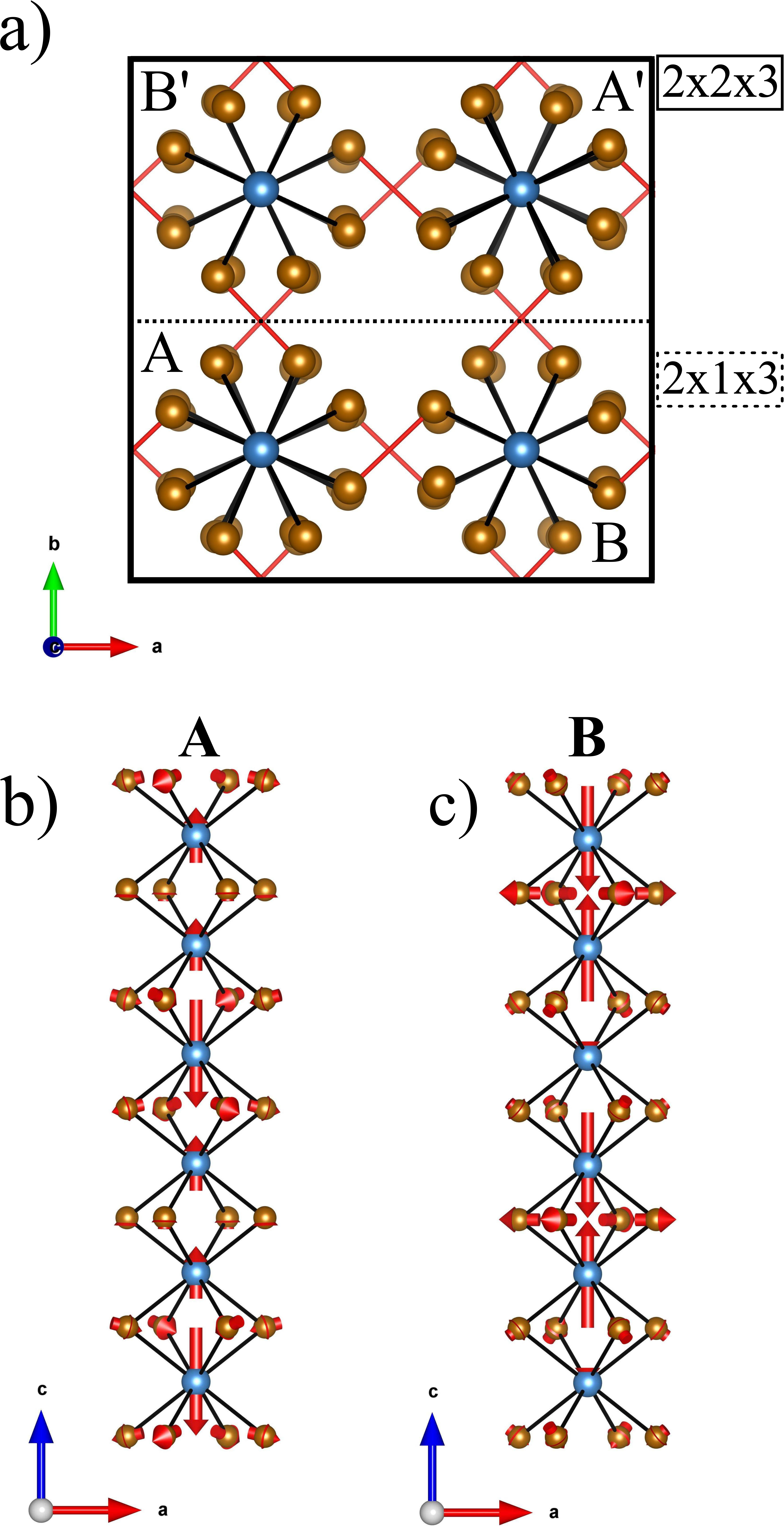}
    \caption{Unstable phonon modes correspondence to each chain type in the 2$\times$2$\times$3 and 2$\times$1$\times$3 TaTe$_4$ supercells.}
     \label{fig:tate4_cdw}
\end{figure*}

\begin{figure*}[!hptb]
    \centering
    \includegraphics[width=0.425\textwidth]{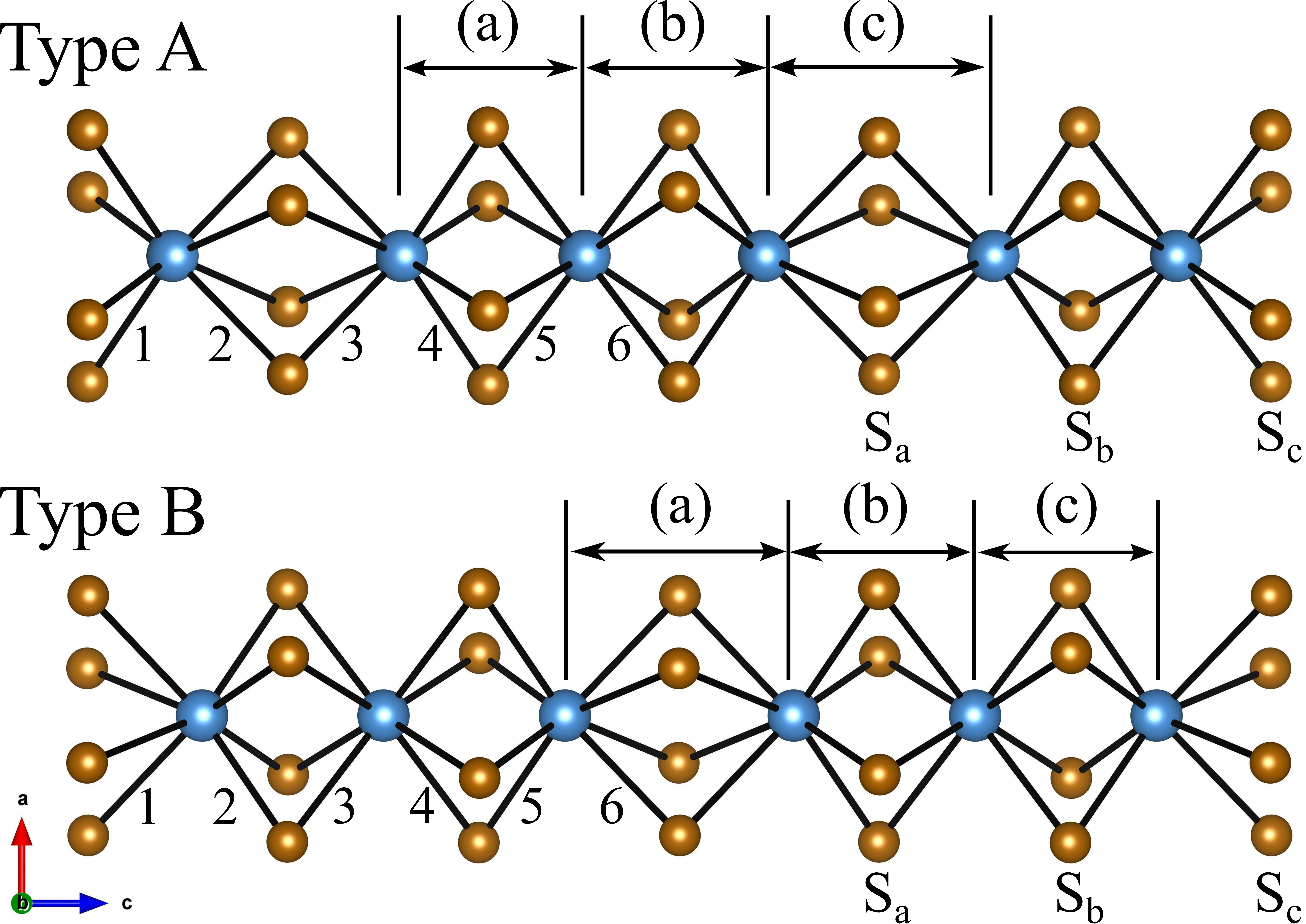}
    \caption{Distances labels in the different modulated chains. Labels 1-6 refer to Ta-Te distances as described in Tab. \ref{table:TaTe}, Ta-Ta distances labeled as (a)-(b) described in Tab. \ref{table:TaTa}, respectively labels S$_a$-S$_c$ corresponding to Te-Te contacts as described in Tab. \ref{table:TeTe}.}
    \label{fig:tate4_distance_labels}
\end{figure*}

The calculated 2$a$\xone2$a$\xone3$c$ phase as well as the 2$a$\xone1$a$\xone3$c$ one can be described on the basis of Fig.~\ref{fig:tate4_cdw}. The unit cell of both phases contains two different TaTe$_4$ chains with different trimerization patterns labelled A and B. In chain A (Fig.~\ref{fig:tate4_cdw}b) three different Ta-Ta distances are generated (see Table~\ref{table:TaTa}); two are shorter than in the average structure and one is longer whereas in chain B (Fig.~\ref{fig:tate4_cdw}c) there are only two different distances; the two shorter Ta-Ta distances are identical in that case but different in chain A. The average value of the two shorter distances in chain A (3.128 \AA) is very similar to the short distance in chain B (3.133 \AA). Trimerized units or clusters are thus formed in every chain increasing the metal-metal bonding. 
Interestingly, the atomic displacements corresponding to these two types of chains can be inferred from linear combinations of the soft-phonon modes obtained for q$_1$=($\frac{1}{2}$,$\frac{1}{2}$,$\frac{1}{3}$) and q$_2$=($\frac{1}{2}$,-$\frac{1}{2}$,$\frac{1}{3}$). In particular, q$_1$+q$_2$ gives the distortion pattern for chain B, while q$_1$-q$_2$ gives that of chain A. This mixing of modes also explains why there are two different types of chains coming from only one soft-phonon branch in the vibrational band structure.
We can infer from these calculations that the charge density wave modulation of TaTe$_4$ is electron-phonon coupling based.

\begin{figure*}[!hptb]
    \centering
    \includegraphics[width=0.975\textwidth]{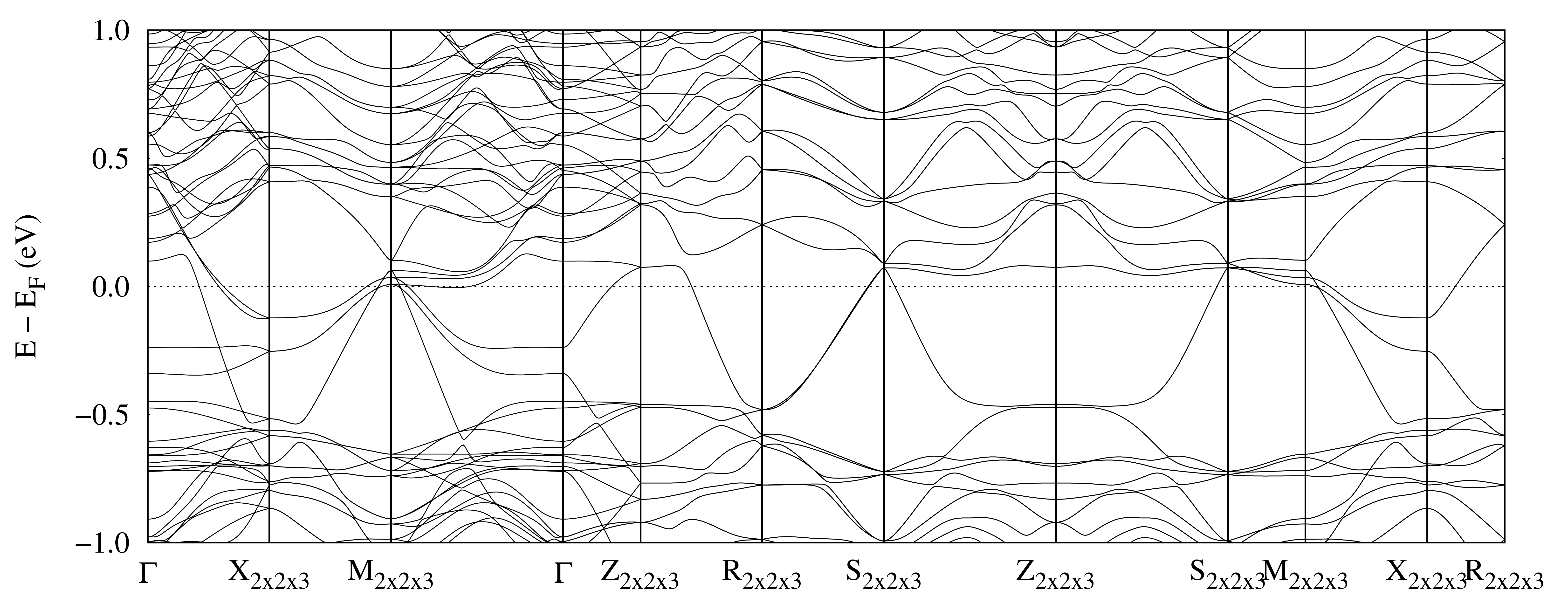}
    \protect\caption{Band structure for the 2$a$\xone2$a$\xone3$c$ modulated TaTe$_4$ structure.}
    \label{fig:tate4_bands_cdw}
\end{figure*}

\begin{figure*}[!hptb]
    \centering
    \includegraphics[width=0.90\textwidth]{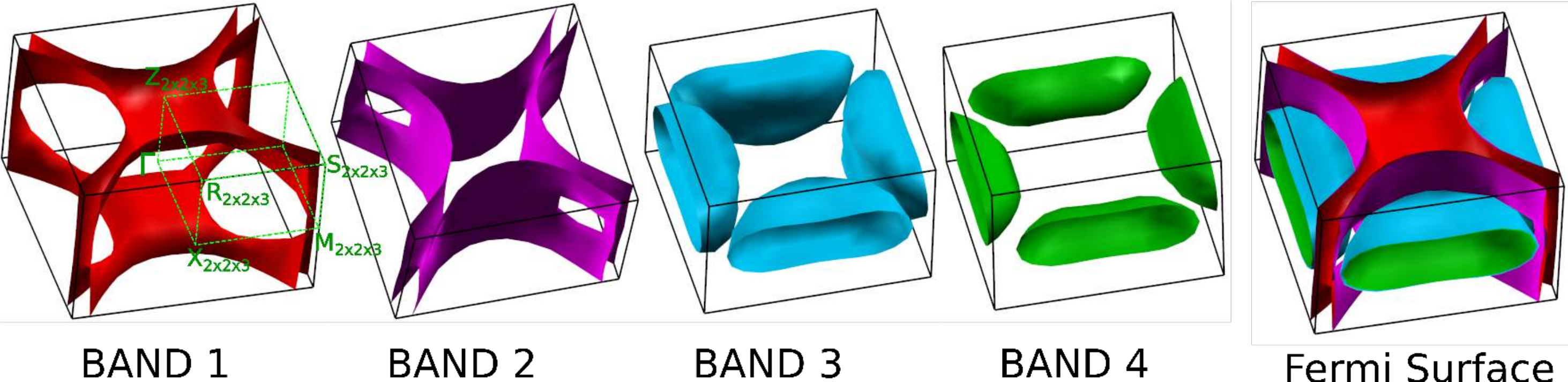}
    \protect\caption{Calculated Fermi surface for the 2$a$\xone2$a$\xone3$c$ modulated TaTe$_4$ structure.}
    \label{fig:tate4_fs_cdw}
\end{figure*}

Although as shown in Figs.~\ref{fig:tate4_cdw}b and c the structural variations are strongly located on the Ta sublattice, the occurrence of these triplets induces a response of the Te sublattice. There are two different types of Te squares (they are really squares in chain A but rectangles in chain B): one-third of them have shorter Te-Te contacts whereas two-thirds have longer Te...Te contacts (see Table~\ref{table:TeTe}). Those being contracted are those in between Ta atoms which have moved apart; this is easy to understand since the TaTe$_4$ chains tend to keep the Ta-Te bonding. The variation of these Te-Te distances is considerably larger (one order of magnitude) than the variation of the Te-Te interchain bonds (see Table~\ref{table:TeTe}). Thus, it is clear that the system tries to keep the Te-Te bonds as stable as possible. The rotations and shifts undergone by the Te atoms under the modulation have this purpose. The variation of the Ta-Te bonds (Table~\ref{table:TaTe}) can be analyzed in the same way. All structural variations noted for the optimized structures mirror those of the experimental ones. In summary it appears that optimization of the metal-metal bonding subject to inducing the minimum alteration of the Te sublattice is the driving force for the CDW irrespective of the nature of the Fermi surface. 

\subsection{C. Electronic structure}\label{sec:Elect_Struct_CDW}

Using this structure we have calculated the band structure of the CDW phase. As it was also found experimentally, the modulated phase is clearly metallic as one can see in Fig. \ref{fig:tate4_bands_cdw}. The extensive folding leads to the four components of the Fermi surface shown in Fig.~\ref{fig:tate4_fs_cdw}. It is clear from this Fermi surface that TaTe$_4$ in the CDW structure is a 3D metal, in agreement with the experimental observation of $\rho_c/\rho_a\sim$ 1~\cite{Tadaki1990,Gao2017,Luo2017,Yang2018,Lima2019}. Since TaTe$_4$ is already in the modulated phase at room temperature it is not possible to know the resistivity change when the transition occurs. However, for NbTe$_4$ which is structurally and electronically very similar, the change in the resistivity is very small without variation of the anisotropy~\cite{Tadaki1990}. This is also consistent with the 3D character of the FS calculated for the average and modulated structures of TaTe$_4$. However, the general shape of the FS has been strongly affected by the modulation. The different components of the Fermi surface in Fig.~\ref{fig:tate4_fs_cdw} contain numerous closed circuits which are consistent with recent studies reporting a very large magnetoresistance for TaTe$_4$ in the CDW state~\cite{Gao2017,Luo2017,Yang2018,Lima2019}. We note the occurrence of a crossing between a steeply raising band with almost linear dispersion and another less dispersive band at the Fermi level near the M point (see the M to X direction in Fig.~\ref{fig:tate4_fs_cdw}). Inclusion of spin-orbit coupling does not alter this situation. The two bands have mixed Ta and Te character and the crossing is a consequence of the structural modulation. This feature is reminiscent of the observation of very large linear magnetoresistance for materials exhibiting a Dirac cone at the Fermi level.~\cite{Novak2015,Kumar2021} and could be at the origin of very large magnetoresistance exhibited by TaTe$_4$. 

Fig.~\ref{fig:tate4_bands_cdw} is in strong contrast with the usual shape of the band structure for metallic quasi-1D transition metal chalcogenides undergoing a structural modulation along the chain direction. As mentioned before, the crystal structure of VS$_4$ although strongly related with those of TaTe$_4$ contains VS$_4$ chains which are the result of a dimerization within the chains (see Fig. S2 in SI). This may be conceptually described as being the result of a usual Peierls dimerization of a chain of transition metal atoms (M) possessing a half-filled quasi-1D $d_{z^{2}}$ band. Note that in the present case since the repeat unit of the average cell already contains two Ta atoms because of the antiprismatic nature of the coordination the distortion is formally a tetramerization. The band structure for such a system should exhibit a semiconducting gap separating two sets of transition metal based bands associated with bonding and antibonding M-M levels resulting from the modulation. This is exactly what is observed in the case of the dimerized structure of VS$_4$ (see Fig. S3 in SI). Both TaTe$_4$ and VS$_4$ contain two X$_2^{2-}$ (X= S or Te) chalcogen-chalcogen bonds (although of different nature) so that the formal electron counting for the transition metal atom is the same in both compounds. However, the chalcogen-chalcogen bonds occur within the chains in VS$_4$ but between the chains in TaTe$_4$. This subtle structural difference completely changes the nature of the electronic structure and leads to a completely different low-temperature structural modulation and physical behaviour.     

\section{Conclusions}\label{sec:conclusions}

We have discussed the electronic structure of TaTe$_4$ and argued that, despite the occurrence of TaTe$_4$ square antiprismatic chains in the crystal structure, this solid has a 3D electronic structure. This is primarily the result of the inter-chain coupling between adjacent chains through Te atoms. The Te-Te inter-chain distances in this compound are compatible with a Te-Te bond. Based on our FS and electron-hole response function calculations we have dismissed a Fermi surface driven instability as the origin of the modulated structure. However, this mechanism could be invoked to explain the modulation of the related VS$_4$ solid which does not exhibit strong inter-chain contacts. We have calculated the phonon dispersion and described the ordering of the possible modulated phases. The optimized 2$a$\xone2$a$\xone3$c$ structure is found to be the more stable in agreement with the experimental observations and it can be obtained directly from a soft-phonon mode computed for the undistorted structure. The nature of the distortions with respect to the average structure suggest that the driving force for the distortion is the maximization of Ta-Ta metal-metal bonding subject to inducing the minimum decrease in bonding within the Te sublatice. 

\section*{Acknowledgements}

This work was supported by Spanish MINECO (the Severo Ochoa Centers of Excellence Program under Grants SEV-2017-0706 and FUNFUTURE CEX2019-000917-S), Spanish MICIU, AEI and EU FEDER (Grants No. PGC2018-096955-B-C43 and No. PGC2018-096955-B-C44), Generalitat de Catalunya (Grant No. 2017SGR1506 and the CERCA Programme), and the European Union  MaX Center of Excellence (EU-H2020 Grant No. 824143). Computational resources have been provided partially by the supercomputing facilities of the Universit\'e catholique de Louvain (CISM/UCL) and the Consortium des \'Equipements de Calcul Intensif en F\'ed\'eration Wallonie Bruxelles (C\'ECI) funded by the Fond de la Recherche Scientifique de Belgique (F.R.S.-FNRS) under convention 2.5020.11 and by the Walloon Region.

\pagebreak
\widetext
\begin{center}
\textbf{\large Supplementary Information for "Competition between Ta-Ta and Te-Te bonding leading to the commensurate charge density wave in TaTe$_4$"}
\end{center}

\setcounter{equation}{0}
\setcounter{figure}{0}
\setcounter{table}{0}
\setcounter{page}{1}
\makeatletter
\renewcommand{\theequation}{S\arabic{equation}}
\renewcommand{\thefigure}{S\arabic{figure}}
\renewcommand{\thetable}{S-\Roman{table}}
\renewcommand{\bibnumfmt}[1]{[S#1]}
\renewcommand{\citenumfont}[1]{S#1}

\author{Bogdan Guster}
\affiliation{Catalan Institute of Nanoscience and Nanotechnology (ICN2), CSIC and The Barcelona Institute of Science and Technology, Campus Bellaterra, 08193 Barcelona, Spain}
\affiliation{European Theoretical Spectroscopy Facility, Institute of Condensed Matter and Nanosciences, Universit\'{e} catholique de Louvain, Chemin des \'{e}toiles 8, bte L07.03.01, B-1348 Louvain-la-Neuve, Belgium}

\author{Miguel Pruneda}
\affiliation{Catalan Institute of Nanoscience and Nanotechnology (ICN2), CSIC and The Barcelona Institute of Science and Technology, Campus Bellaterra, 08193 Barcelona, Spain}

\author{Pablo Ordej\'on}
\affiliation{Catalan Institute of Nanoscience and Nanotechnology (ICN2), CSIC and The Barcelona Institute of Science and Technology, Campus Bellaterra, 08193 Barcelona, Spain}

\author{Enric Canadell}
\affiliation{Institut de Ci\`encia de Materials de Barcelona (ICMAB-CSIC), Campus UAB, 08193 Bellaterra, Spain}

\maketitle

\vspace{50pt}

\begin{center}
    \textbf{\LARGE Contents}
\end{center}

\noindent \textbf{Tables}

\noindent
\begin{itemize}
    \item[S-I.]{{Comparison between the experimental coordinates and those calculated for the TaTa$_4$ modulated structure in its 2$a$\xone2$a$\xone3$c$ phase.}}
\end{itemize}

\noindent \textbf{Figures}

\noindent
\begin{itemize}
    \item[S1.]{{View of one of the tellurium ($aa$) planes in the crystal structure of TaTe$_4$ showing in-plane Te...Te short contacts and Te-Te bonds. The distances refer to the average crystal structure~\cite{Bronsema1987}.}}
\end{itemize}

\noindent
\begin{itemize}
    \item[S2.]{{(a) Top view of the VS$_4$ crystal structure. (b) Lateral view of a VS$_4$ chain showing the rectangular antiprismatic coordination of the V atoms and the structural dimerization (i.e. four V atoms per repeat unit). The full and dashed lines refer to the short and long S-S contacts in the rectangular S$_4$ units~\cite{Allman1964}.}}
\end{itemize}

\noindent
\begin{itemize}
    \item[S3.]{{Fatband structure of VS$_4$ where the green and purple full circles are associated with the S 3$p$ and V 3$d_{z^2}$ contributions.}}
\end{itemize}

\noindent
\begin{itemize}
    \item[S4.]{{Comparison between band structures with and without the inclusion of Spin-Orbit (SO) coupling in the 2$a$\xone2$a$\xone3$c$ modulated TaTe$_4$ structure.}}
\end{itemize}

\clearpage

\begin{table}[!hptb]
\centering
\caption{Comparison between the experimental coordinates (taken from \cite{Bronsema1987}) and those calculated for the TaTe$_4$ modulated structure in its 2$a$\xone2$a$\xone3$c$ phase.\\\\}
\begin{tabular}{|c|c|c|c|c|c|c|c|}
\cline{3-8}
\multicolumn{2}{c}{} & \multicolumn{3}{|c|}{Experimental} & \multicolumn{3}{c|}{Calculated} \\ \hline
\multicolumn{1}{|c|}{\begin{tabular}[c]{@{}c@{}}Atom\\ type\end{tabular}} & \begin{tabular}[c]{@{}c@{}}Wyckoff\\ Position\end{tabular} & x & y & z & x & y & z \\ \hline
\multicolumn{1}{|c|}{Ta 1} & 8e & 0.0000 & 0.0000 & 0.09652 & 0.0000 & 0.0000 & 0.09668 \\ \hline
\multicolumn{1}{|c|}{Ta 2} & 4a & 0.0000 & 0.0000 & 0.25000 & 0.0000 & 0.000 & 0.25000 \\ \hline
\multicolumn{1}{|c|}{Ta 3} & 4c & 0.0000 & 0.5000 & 0.41827 & 0.0000 & 0.5000 & 0.41828 \\ \hline
\multicolumn{1}{|c|}{Ta 4} & 4c & 0.0000 & 0.5000 & 0.26351 & 0.0000 & 0.5000 & 0.26385 \\ \hline
\multicolumn{1}{|c|}{Ta 5} & 4c & 0.0000 & 0.5000 & 0.07030 & 0.0000 & 0.5000 & 0.07005 \\ \hline
\multicolumn{1}{|c|}{Te 1} & 16g & 0.0611 & 0.1579 & -0.00300 & 0.0832 & 0.1653 & -0.00176 \\ \hline
\multicolumn{1}{|c|}{Te 2} & 16g & 0.5802 & 0.1655 & 0.00170 & 0.5911 & 0.1874 & -0.00049 \\ \hline
\multicolumn{1}{|c|}{Te 3} & 16g & 0.1656 & 0.0789 & 0.16860 & 0.1828 & 0.0933 & 0.16734 \\ \hline
\multicolumn{1}{|c|}{Te 4} & 16g & 0.6665 & 0.5747 & 0.16970 & 0.6774 & 0.5796 & 0.16336 \\ \hline
\multicolumn{1}{|c|}{Te 5} & 16g & 0.6687 & 0.0700 & 0.16330 & 0.6673 & 0.0827 & 0.16871 \\ \hline
\multicolumn{1}{|c|}{Te 6} & 16g & 0.1561 & 0.5648 & 0.16750 & 0.1717 & 0.5817 & 0.16943 \\ \hline
\end{tabular}
\label{table:tate4_structure_cdw}
\end{table}

\clearpage

\begin{figure}[!hptb]
    \centering
    \includegraphics[width=0.45\textwidth]{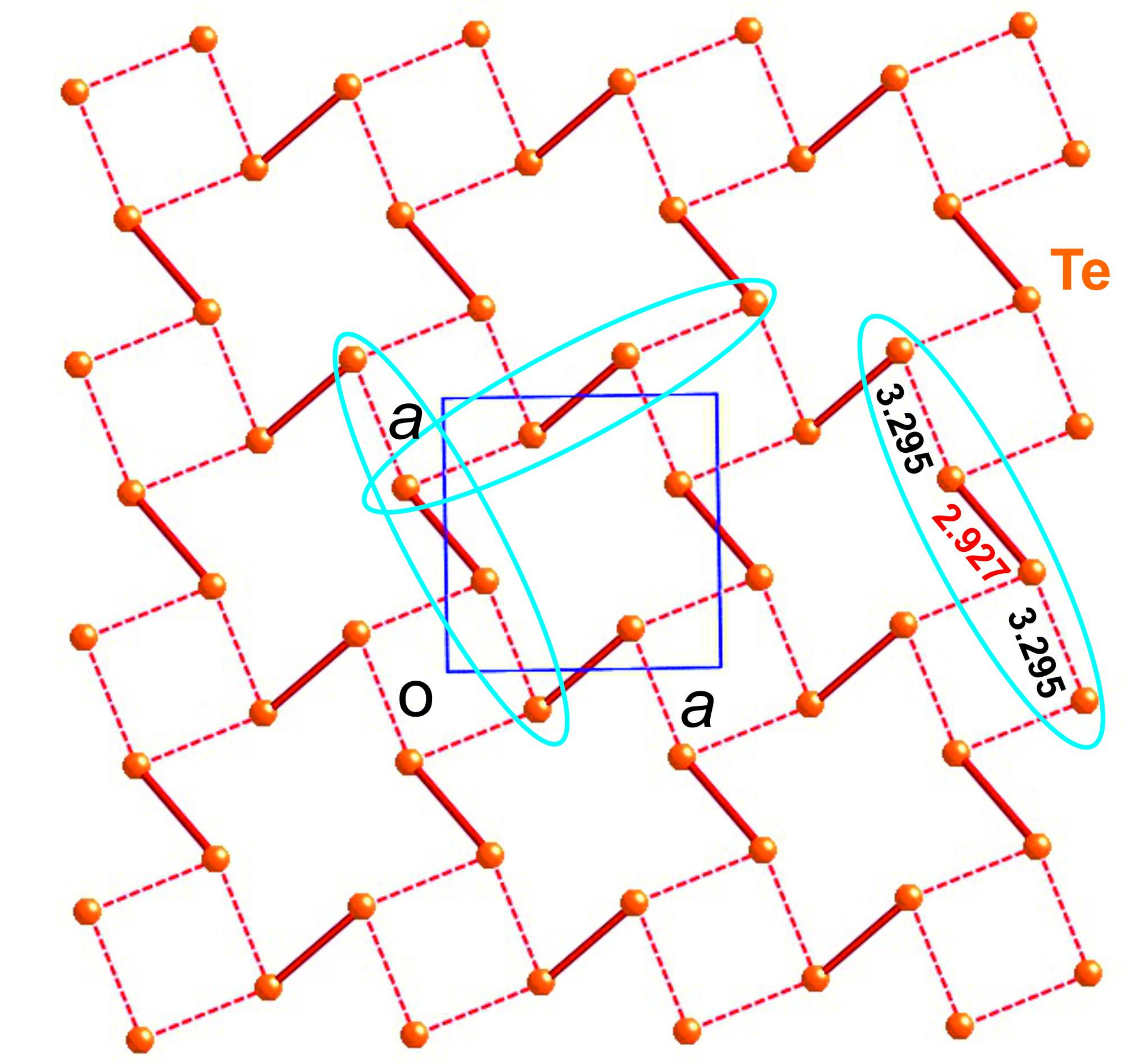}
    \caption{{View of one of the tellurium ($aa$) planes in the crystal structure of TaTe$_4$ showing in-plane Te...Te short contacts and Te-Te bonds. The distances refer to the average crystal structure~\cite{Bronsema1987}.}}
    \label{fig:te-te_plane}
\end{figure}

\clearpage

\begin{figure}[!hptb]
    \centering
    \includegraphics[width=0.45\textwidth]{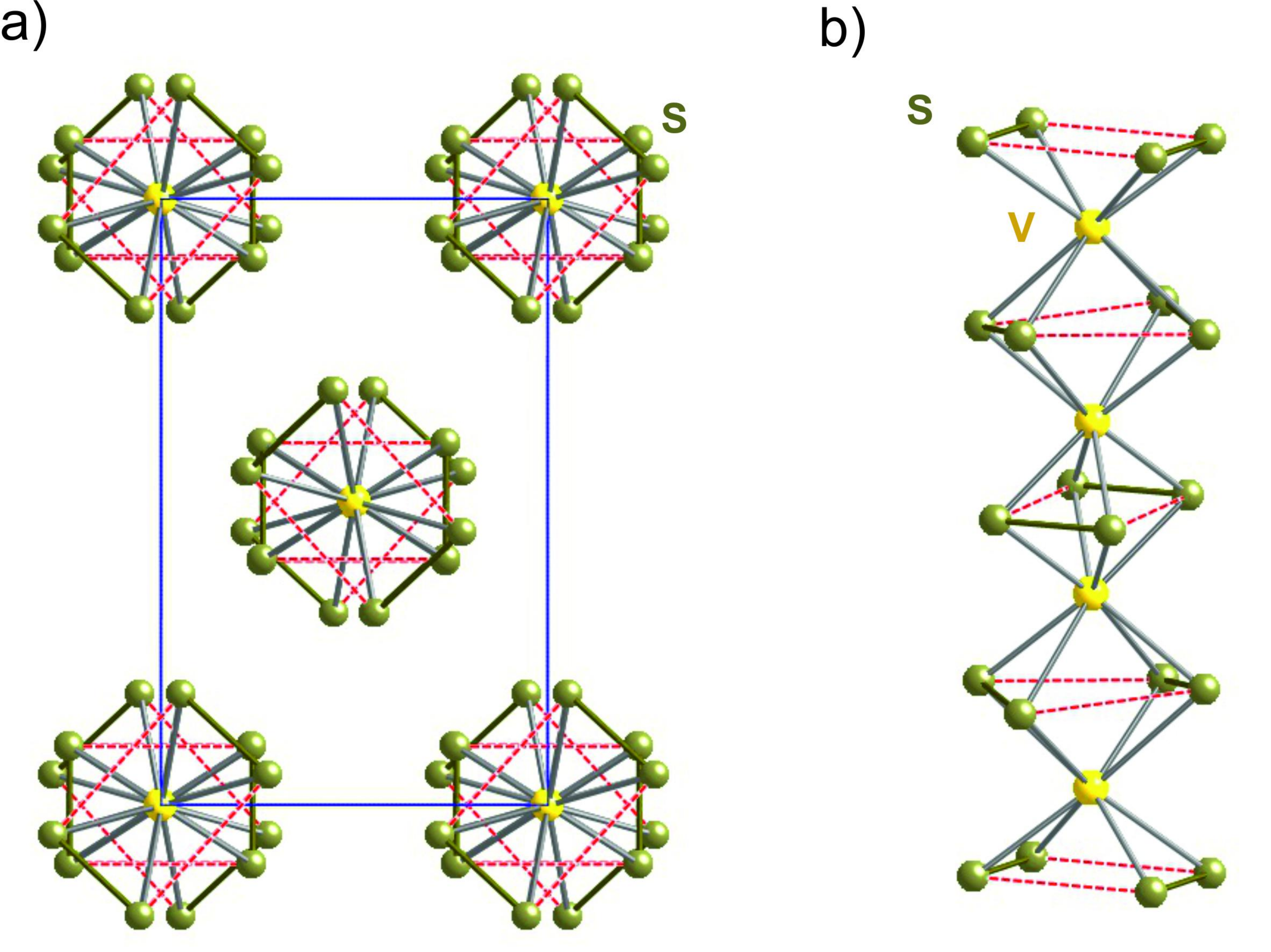}
    \caption{{(a) Top view of the VS$_4$ crystal structure. (b) Lateral view of a VS$_4$ chain showing the rectangular antiprismatic coordination of the V atoms and the structural dimerization (i.e. four V atoms per repeat unit). The full and dashed lines refer to the short and long S-S contacts in the rectangular S$_4$ units~\cite{Allman1964}.}}
    \label{fig:vs4_structure}
\end{figure}

\begin{figure}[!hptb]
    \centering
    \includegraphics[width=0.5\textwidth]{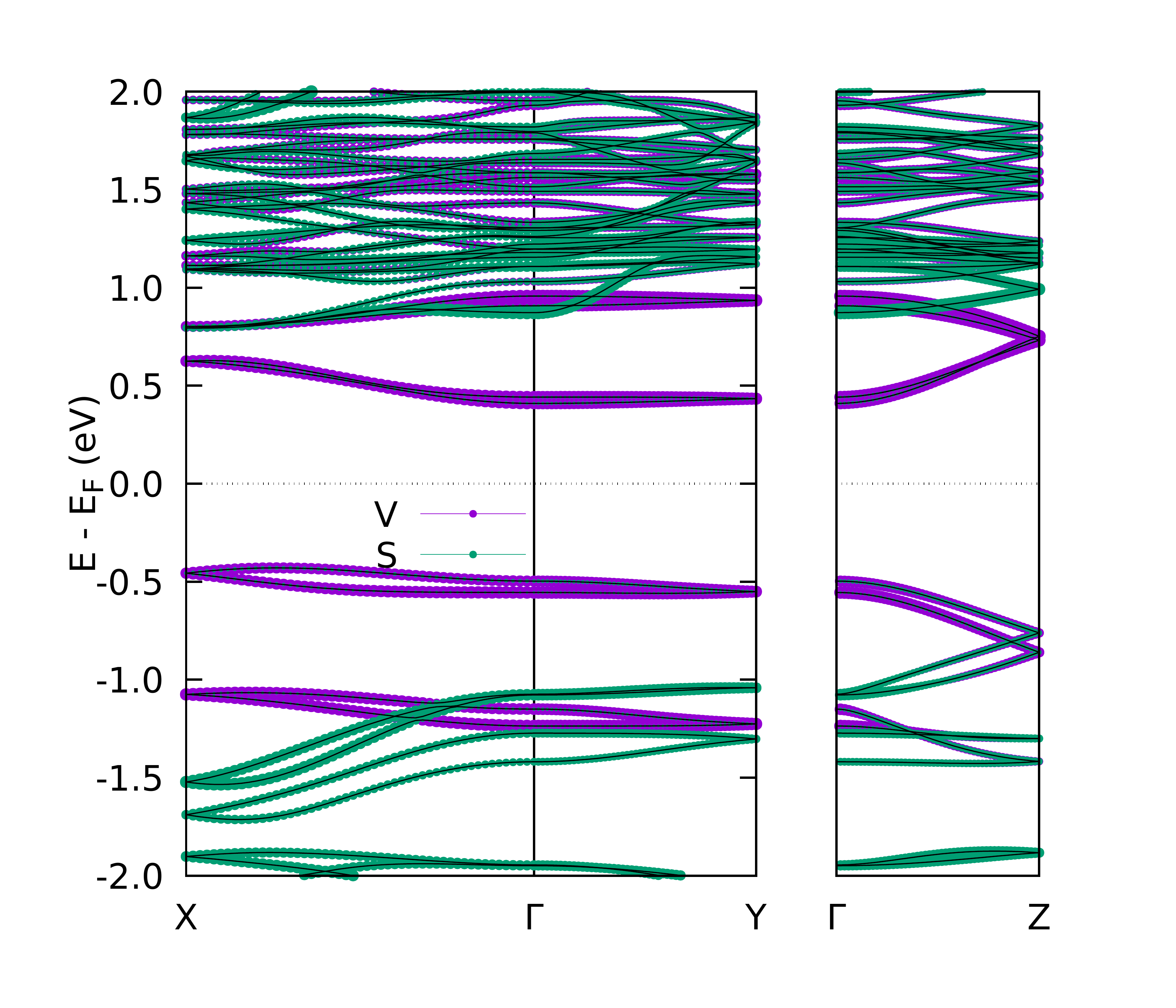}
    \caption{{Fatband structure of VS$_4$ where the green and purple full circles are associated with the S 3$p$ and V 3$d_{z^2}$ contributions.}}
    \label{fig:te-te_plane}
\end{figure}

\begin{figure}[!hptb]
    \centering
    \includegraphics[width=0.95\textwidth]{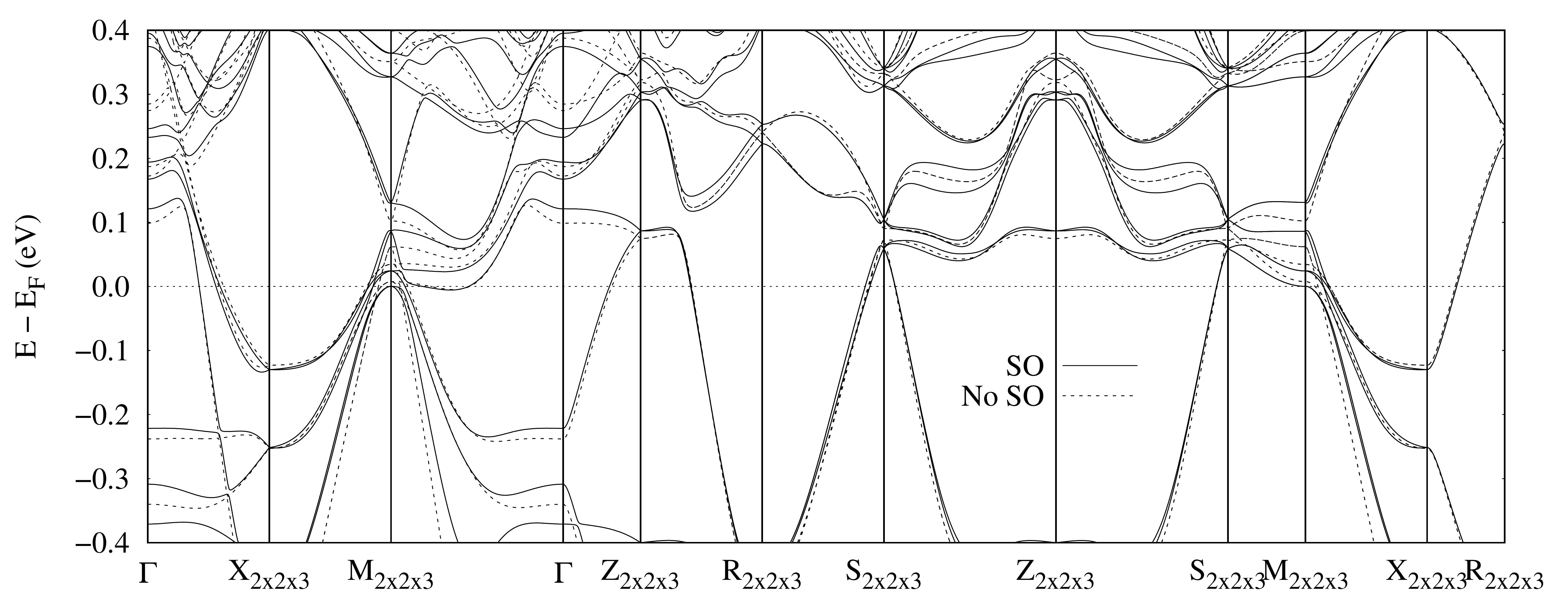}
    \caption{{Comparison between band structures with and without the inclusion of Spin-Orbit (SO) coupling in the 2$a$\xone2$a$\xone3$c$ modulated TaTe$_4$ structure.}}
    \label{fig:te-te_plane}
\end{figure}

\end{document}